\documentclass[a4paper]{article}
\usepackage[german, english]{babel}
\usepackage{a4wide}
\usepackage{amsmath}
\usepackage{amssymb}
\usepackage{ifthen}
\usepackage{graphicx}
\usepackage{epsfig}
\usepackage[hang,nooneline]{subfigure}
\newcounter{fig}

\begin{document}

\title{\bf Scalarized Hairy Black Holes}
\vspace{1.5truecm}
\author{
{\bf Burkhard Kleihaus$^a$, Jutta Kunz$^a$
  and Stoytcho Yazadjiev$^b$}\\[5pt]
$^a$Institut f\"ur  Physik, Universit\"at Oldenburg, Postfach 2503\\
  D-26111 Oldenburg, Germany\\
$^b$Department of Theoretical Physics, Faculty of Physics, Sofia University\\
  Sofia 1164, Bulgaria\\
}

\vspace{1.5truecm}

\date{\today}

\maketitle
\vspace{1.0truecm}

\begin{abstract}
In the presence of a complex scalar field
scalar-tensor theory allows 
for scalarized rotating hairy black holes. 
We exhibit the domain of existence for these scalarized black holes,
which is bounded by scalarized rotating boson stars and 
ordinary hairy black holes.
We discuss the global properties of these solutions.
Like their counterparts in general relativity,
their angular momentum may exceed the Kerr bound,
and their ergosurfaces may consist of a sphere and a ring,
i.e., form an ergo-Saturn.
\end{abstract}

\section{Introduction}

One of the major discoveries in physics during the last two decades  
was the accelerated expansion of the Universe. 
General relativity and the standard model of particle physics 
fail to explain this phenomenon. 
This situation calls for new alternative ideas 
able to give a satisfactory explanation of the cosmological observations. 
One of the possibilities is to go beyond general relativity 
and to consider more general theories of gravity. 

Among the most natural generalizations of the original Einstein theory 
are the scalar-tensor theories
\cite{Jordan:1949zz,Fierz:1956zz,Jordan:1959eg,Brans:1961sx,Dicke:1961gz}.

These theories are viable gravitational theories 
and can pass all known experimental and observational constraints.
In addition, they can explain the accelerated expansion of the Universe. 
The scalar-tensor generalizations of the original Einstein theory 
naturally arise in the context of the modern unifying theories 
as string theory and Kaluza-Klein theories. 

In scalar-tensor theories the gravitational interaction 
is mediated not only by the spacetime metric 
but also by an additional scalar field. 
From a physical point of view this scalar field plays 
the role of a variable gravitational constant.

General relativity (GR) is well-tested in the weak-field regime, 
whereas the strong-field regime remains largely unexplored and unconstrained.
In the strong-field regime one expects the differences 
between GR and alternative theories of gravity to be more pronounced. 
The natural laboratories for testing the strong-field regime 
of gravitational theories are compact stars and black holes.

There exist scalar-tensor theories which are indistinguishable from GR 
in the weak-field regime but which can differ significantly from GR 
in the strong-field regime. 
An example of such a phenomenon is the so-called spontaneous scalarization,
observed in a certain class of scalar-tensor theories. 
When spontaneous scalarization takes place, in addition to
the general relativistic solutions with a trivial scalar field, 
there exist further solutions with a nontrivial scalar field.
In fact, these scalarized solutions are energetically more favorable 
than their GR counterparts. 

Spontaneous scalarization was first observed for neutron stars
\cite{Damour:1993hw}, where
{\sl spectacular changes} were seen in static equilibrium configurations 
for a given nuclear equation of state.
More recently, spontaneous scalarization was also observed in
rapidly rotating neutron stars \cite{Doneva:2013qva,Doneva:2014uma},
where the deviations of the rapidly rotating scalar-tensor neutron stars 
from the general-relativistic solutions were even
significantly larger than in the static case.

Spontaneous scalarization was also observed for static uncharged 
and charged boson stars \cite{Whinnett:1999sc,Alcubierre:2010ea,Ruiz:2012jt}.
The first purpose of the present paper is to study 
rapidly rotating boson stars in scalar-tensor theories,
and to establish the phenomenon of spontaneous scalarizarion 
for these stationary compact objects.
The second purpose of this paper is to address the existence
of scalarized hairy black holes.

In General Relativity (GR) rotating vacuum black holes are 
described in terms of the Kerr solution.
This solution specifies the full spacetime in terms of
only two parameters, its mass and its angular momentum.
Hairy black holes appear, when suitable matter fields are included.
Examples are chiral fields, Yang-Mills and Higgs fields,
yielding hairy static black holes
\cite{Luckock:1986tr,Volkov:1989fi,Breitenlohner:1991aa,Kleihaus:1997ws}
as well as rapidly rotating hairy black holes
\cite{Kleihaus:2000kg,Kleihaus:2004gm}.

Recently it was noted, that also a single complex scalar field
allows for hairy black holes, provided the black holes are
rotating \cite{Herdeiro:2014goa}.
In fact, these solutions maybe viewed as a generalization of
rotating boson stars, that are endowed with a horizon.
The regular boson stars form part of the boundary of the
domain of existence of this new type of hairy black holes.
The other parts of the boundary exist of extremal hairy black holes
and scalar clouds.

Here we show, that besides these rapidly rotating hairy black holes,
already present in GR, 
scalar-tensor theory again allows for the phenomenon of scalarization.
In particular,
we study the physical properties of these scalarized hairy black holes,
and map their domain of existence.

\section{Scalar-Tensor Theories}

Denoting the gravitational scalar by $\Phi$,  
the gravitational action of scalar-tensor theories
in the physical Jordan frame is given by
\begin{eqnarray} \label{JFA}
S = {1\over 16\pi G_{*}} \int d^4x \sqrt{-{\tilde
g}}\left({F(\Phi)\tilde R} - Z(\Phi){\tilde
g}^{\mu\nu}\partial_{\mu}\Phi
\partial_{\nu}\Phi   -2 W(\Phi) \right) +
S_{m}\left[\Psi_{m};{\tilde g}_{\mu\nu}\right] ,
\end{eqnarray}
where $G_{*}$ is the bare gravitational constant,
${\tilde g}_{\mu\nu}$ is the spacetime metric, ${\tilde R}$ is
the Ricci scalar curvature, 
and $S_{m}\left[\Psi_{m};{\tilde g}_{\mu\nu}\right]$ denotes the
action of the matter fields.

The functions $F(\Phi)$, $Z(\Phi)$ and $W(\Phi)$
are subject to physical restrictions:
We require $F(\Phi)>0$, since gravitons should carry positive energy,
and $2F(\Phi)Z(\Phi) + 3[dF(\Phi)/d\Phi]^2 \ge 0$,
since the kinetic energy of the saclar field should not be negative.
The matter action $S_{m}$ depends on the matter field $\Psi_{m}$ 
and on the space-time metric ${\tilde g}_{\mu\nu}$.
The matter action does not involve the gravitational scalar field $\Phi$
in order to satisfy the weak equivalence principle.

Variation of the action with respect to the spacetime metric and
the gravitational scalar as well as the matter field
leads to the field equations in the Jordan frame.
However, these field equations are rather involved.
It is therefore easier to consider a mathematically equivalent
formulation of scalar-tensor theories 
in the conformally related Einstein frame
with metric $g_{\mu\nu}$
\begin{equation}\label {CONF1}
g_{\mu\nu} = F(\Phi){\tilde g}_{\mu\nu} .
\end{equation}

In the Einstein frame the action then becomes (up to a boundary term)
\begin{eqnarray}
S= {1\over 16\pi G_{*}}\int d^4x \sqrt{-g} \left(R -
2g^{\mu\nu}\partial_{\mu}\varphi \partial_{\nu}\varphi -
4V(\varphi)\right)+ S_{m}[\Psi_{m}; {\cal A}^{2}(\varphi)g_{\mu\nu}] ,
\end{eqnarray}
where $R$ is the Ricci scalar curvature with respect to the Einstein
metric $g_{\mu\nu}$,
$\varphi$ represents the new scalar field defined via
\begin{equation}\label {CONF2}
\left(d\varphi \over d\Phi \right)^2 = {3\over
4}\left({d\ln(F(\Phi))\over d\Phi } \right)^2 + {Z(\Phi)\over 2
F(\Phi)}
\end{equation}
with the new functions
\begin{equation}\label{CONF3}
{\cal A}(\varphi) = F^{-1/2}(\Phi) \,\,\, ,\nonumber \\
2V(\varphi) = W(\Phi)F^{-2}(\Phi).
\end{equation}

By varying this action with respect to the metric in the
Einstein frame $g_{\mu\nu}$, the scalar field $\varphi$,
and the matter field $\Psi_m$,
we find the set of field equations in the Einstein frame.

In particular, we here consider a
scalar tensor theory with potential $V(\varphi)=0$
and function ${\cal A}(\varphi)$ \cite{Damour:1993hw}
\begin{equation}
\ln {\cal A}(\varphi) = \frac{1}{2} \beta \varphi^2 \ .
\label{funA}
\end{equation}
We present a systematic study for the parameter value
$\beta=-4.7$, but we have also done calculations for
larger values of $\beta$.

For the matter action we choose a complex boson field $\Psi$
\begin{equation}\label{Smat}
S_{m}\left[\Psi_{m};{\tilde g}_{\mu\nu}\right]
= - \int d^4x \sqrt{-g} 
\left[ \frac{1}{2} {\cal A}^2(\varphi)  g^{\mu\nu}
\left( \Psi_{, \, \mu}^* \Psi_{, \, \nu} 
+ \Psi _ {, \, \nu}^* \Psi _{, \, \mu} \right) 
+ {\cal A}^4(\varphi)  U( \left| \Psi \right|) \right]
\end{equation}
with self-interaction potential
\begin{equation}\label{SmatU}
U( \left| \Psi \right|) =
m_b^2 \left| \Psi \right|^2 + \Lambda \left| \Psi \right|^4 \ .
\end{equation}

\section{Properties}

To obtain stationary hairy black hole and boson star solutions
we employ the line element
\begin{equation}
ds^2 = - f_0 N dt^2 + \frac{f_1}{f_0} \left( f_2 \left[
\frac{1}{N} d\bar{r}^2 + \bar{r}^2 d \theta^2 \right]
+ \bar{r}^2 \sin^2 \theta \left[ d\varphi - f_3 dt \right]^2 \right) ,
\label{metric}
\end{equation}
with the metric functions $f_i(\bar{r},\theta)$, $i=0,...,3$, and
$$ N = 1 - \frac{\bar{r}_{\rm H}}{\bar r}, $$
where $\bar{r}_H$ denotes the horizon parameter.
For the boson stars we seet $\bar{r}_{\rm H}=0$, i.e. $N=1$.
Likewise, we parametrize the gravitational scalar field $\Phi$ by
$\Phi(\bar{r},\theta)$.

For the boson field $\Psi$ we adopt the stationary Ansatz
\begin{equation}
\Psi = \psi(\bar{r},\theta) \, e^{i \omega t + i n \phi}
\end{equation}
where $\psi$ is a real function,
$\omega$ denotes the boson frequency,
and --
as required by the single-valuedness of the scalar field --
$n$ is an integer representing a rotational quantum number.

In the Lagrangian for the boson field $\Psi$ 
we employ a quartic self-interaction potential (\ref{SmatU})
with coupling constant $\Lambda$.
While we present our results for the value $\Lambda=300$,
we have also performed calculations for other values of $\Lambda$.

As in \cite{Herdeiro:2014goa} we introduce a new radial coordinate
\begin{equation}
r = \sqrt{\bar{r}^2 - \bar{r}_{\rm H}^2}
\end{equation}
such that $0 \le r < \infty$, and the event horizon is located at $r=0$.
The functions satisfy the following set of boundary conditions,
obtained from the requirements of asymptotic flatness
as well as of regularity 
%
at the origin and the event horizon in the case of the boson star, resp.
black hole solutions 
\begin{eqnarray}
& & \partial_r f_i(0,\theta) = 0 , \ \ \ i=0,1,2  \ ,  \ \ \
 \psi(0,\theta) = 0 \ , \partial_r \phi(0,\theta) = 0 \\
& &  f_i(\infty,\theta) = 1, \ \ \ i=0,1,2\ , \ \ \ f_3(\infty,\theta) =0 \ ,  \ \ \
 \psi(\infty,\theta) =\phi(\infty,\theta) = 0 \\
& &  \partial_\theta f_i(r,0) = 0 , \ \ \ i=0,1,3 \ , \ \ \ f_2(r,0) =1 \ ,  \ \ \
 \psi(r,0) = 0 \ , \ \ \partial_\theta \phi(r,0) = 0 \\
& &  \partial_\theta f_i(r,\pi/2) = 0 , \ \ \ i=0,1,2,3 \ ,  \ \ \
 \partial_\theta \psi(r,\pi/2) =  \partial_\theta \phi(r,\pi/2) = 0 \ ,
 \end{eqnarray}
and $\partial_r f_3(0,\theta) = 0$, resp. $f_3(0,\theta) =\Omega_{\rm H}$
for boson star and black hole solutions.

The mass $M$ and the angular momentum $J$
of stationary asymptotically flat space-times
can be obtained
in scalar-tensor theory 
- analogously to GR -
from the asymptotic behavior of the
metric functions $f_0$ and $f_3$, respectively,
\begin{equation} 
f_0 \to 1 -2\mu/r \ , \ \ \ f_3  \to  2 J/r^3 \ , 
\end{equation} 
and
\begin{equation}
M = \mu + \frac{\bar{r}_H}{2}  \ .
\end{equation}

Since the action is invariant under the global phase transformation
$\Psi \rightarrow \Psi e^{i\chi}$, 
a conserved current arises
\begin{eqnarray}
j^{\mu} & = &  - i {\cal A}^2(\varphi)\left( \Psi^* \partial^{\mu} \Psi 
 - \Psi \partial^{\mu}\Psi ^* \right) \ , \ \ \
j^{\mu} _{\ ; \, \mu}  =  0 \ .
\end{eqnarray}
It is associated with the Noether charge $Q$ representing the particle number,
\begin{equation}
Q = \int_{\Sigma} \sqrt{ - g} j^0 
d\bar{r} d\theta d\varphi 
\label{Q}
\end{equation}

At the event horizon of the hairy black holes 
the Killing vector $\chi$
\begin{equation}
\chi = \xi + \Omega_{\rm H} \eta \ 
\label{chi} \end{equation}
is null,
and $\Omega_{\rm H}$ represents the horizon angular velocity.
Hairy black holes satisfy \cite{Herdeiro:2014goa}
\begin{equation}
\omega= n \Omega_{\rm H}
\label{omegas} \end{equation}

We denote the horizon area by $A_{\rm H}$ in the Einstein frame, 
and define the areal horizon radius by 
\begin{equation}
R_{\rm H}=\sqrt{A_{\rm H}/4 \pi} \ .
\label{R_H}
\end{equation}
The horizon temperature by $T_{\rm H}$
is obtained from the surface gravity $\kappa_{\rm sg}$,
\begin{equation}
 \kappa_{\rm sg}^2 = 
\left. -\frac{1}{2}  
      (\nabla_\mu \chi_\nu) (\nabla^\mu \chi^\nu)
                    \right|_{\rm H}  ,
\label{surf_gravity}
\end{equation}
and $T=\kappa_{\rm sg}/2 \pi$.

We note, that the limit $\bar{r}_{\rm H} \to 0$
comprises two different types of configurations\footnote{Recall, 
that an extremal black hole has a vanishing horizon parameter
in the coordinates employed.}:

\vspace{-0.2cm}
\begin{itemize}
\itemsep=-2pt
\item[(i)] Extremal black holes are obtained, when $T_{\rm H} \to 0$,
\item[(ii)] Globally regular solutions are obtained,
when $A_{\rm H} \to 0$.
\end{itemize}
\vspace{-0.2cm}
The latter correspond to rotating boson stars.
For boson stars $J=nQ$.

\section{Results}

We have obtained rapidly rotating scalarized boson stars 
and hairy black hole solutions for
a sequence of rotational quantum numbers $n > 0$.
To this end
we have solved the set of coupled non-linear partial differential equations
subject to the appropriate boundary conditions
with a numerical algorithm based on the Newton-Raphson method
\cite{FIDISOL}.
Compactifying space by introducing the radial coordinate
\begin{equation}
x = \frac{r}{1+r} \ .
\label{rcomp} \end{equation}
we have discretized the equations on a non-equidistant grid in
$x$ and  $\theta$,
with typical grid sizes on the order of $250 \times 65$,
covering the integration region
$0\leq x \leq 1$ and $0\leq\theta\leq\pi/2$.
In the following we will used scaled quantities $M/M_0$, $Q/Q_0$, $\omega/\omega_0$
$J/J_0$, and $R_{\rm H}/r_0$, 
with $M_0 = m_{\rm Pl}^2/m_b$,  $Q_0 = J_0= m_{\rm Pl}^2/m_b^2$, $\omega_0=m_b$, $r_0=1/m_b$,
where $m_{\rm Pl}$ denotes the Plack mass. 

\subsection{Scalarized boson stars}

\begin{figure}[h!]
\begin{center}
\vspace{-0.5cm}
\mbox{
(a)\hspace*{-0.5cm}\includegraphics[height=.225\textheight, angle =0]{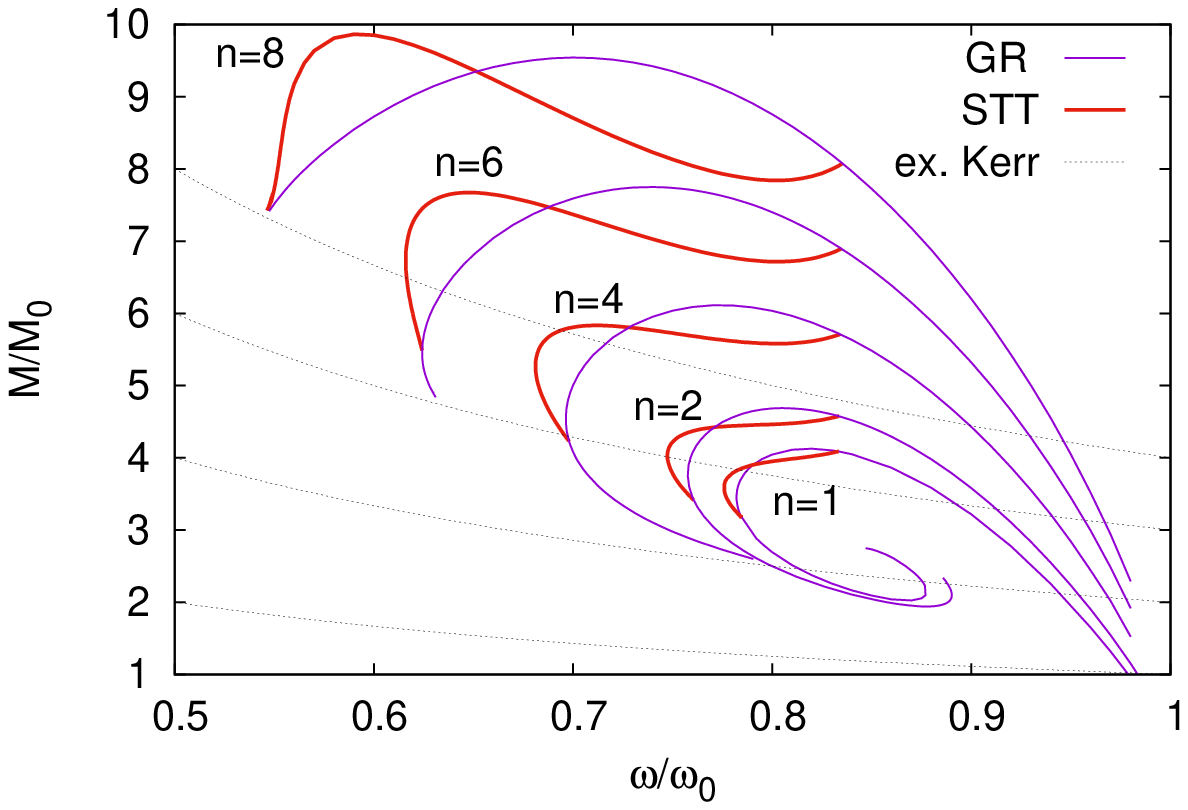}
(b)\hspace*{-0.5cm}\includegraphics[height=.225\textheight, angle =0]{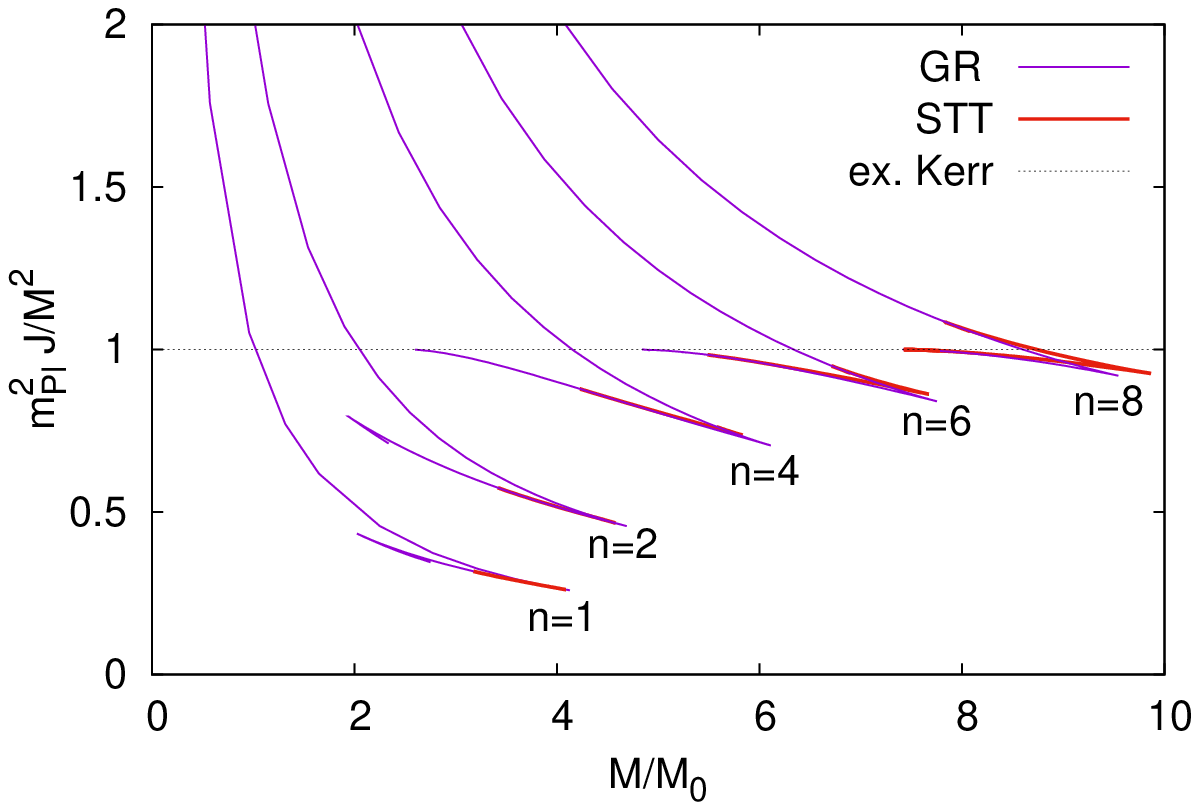}
}
\end{center}
\vspace{-0.5cm}
\caption{
Families of scalarized (thick red)
and ordinary (thin blue) boson star solutions for
rotational quantum number $n=1$, 2, 4, 6 and 8
as well as the corresponding curve(s) for the extremal Kerr solution
(dotted black).
(a) The scaled mass $M/M_0$  versus the scaled
boson frequency $\omega/\omega_0$.
(For black holes $\omega=n\Omega_{\rm H}$.)
(b) The scaled angular momentum $ m_{\rm Pl}^2 J/M^2$ versus the scaled mass $M/M_0$.
\label{fig1}
}
\end{figure}

The globally regular ordinary boson star solutions form a large
part of the boundary of
the domain of existence of the hairy black hole solutions \cite{Herdeiro:2014goa}.
Let us therefore first consider the boson star solutions.
In Fig.~\ref{fig1}(a) the scaled mass $M/M_0$ of the boson star solutions is exhibited 
versus the scaled boson frequency $\omega/\omega_0$
for rotational quantum numbers $n=1$, 2, 4, 6 and 8.
The families of ordinary boson star solutions emerge from the vacuum at 
$\omega=m_b=\omega_0$.
They form a first (and at least in part classically stable) branch, 
until the mass reaches its maximal value.
We note, that this maximal value of the mass increases rapidly with $n$.

Beyond the maximal mass
the families of ordinary boson star solutions continue in a spiral-like manner
for the lowest values of $n$. They end in a merger solution,
where a branch of extremal hairy black hole solutions is encountered
\cite{Herdeiro:2014goa}.
For the higher $n$, however, they feature only a single further branch,
their second branch,
before they merge with a branch of extremal hairy black hole solutions.
Interestingly, each of these higher $n$ second branches of
boson star solutions ends 
close to an extremal Kerr black hole solution, possessing almost
the same mass and the respective horizon angular 
velocity $\Omega_{\rm H}=\omega/n$.

When considering the scaled angular momentum $ m_{\rm Pl}^2 J/M^2$ of these
ordinary boson star solutions versus the mass $M/M_0$ 
as exhibited in Fig.~\ref{fig1}(b)
together with the line $ m_{\rm Pl}^2 J/M^2=1$ of extremal Kerr black holes,
the branches form cusps at extremal values of the mass.
The lower $n$ solutions therefore feature several cusps,
whereas the higher $n$ solutions have a single cusp.
Clearly, for the higher $n$ families of boson star solutions
the second branches approach the extremal Kerr value closely,
when they end in a merger solution,
where a branch of extremal hairy black hole solutions is encountered.


In addition, Fig.~\ref{fig1} exhibits the scalarized boson star solutions 
associated with the ordinary boson star solutions. 
For a given value of $n$ the scalarization arises
at a critical value of the boson frequency, $\omega_{\rm cr}^1$,
where a branch of scalarized boson star solutions emerges 
from the first branch of ordinary boson star solutions.
Interestingly, $\omega_{\rm cr}^1$ is rather independent of 
the rotational quantum number $n$.
The families of scalarized boson stars then extend up to
a second critical value $\omega_{\rm cr}^2$, 
where they merge again into the 
respective second branch of ordinary boson stars.
Since $\omega_{\rm cr}^2$ decreases with $n$,
the domain of existence of rapidly rotating scalarized boson stars
increases with $n$.

For $n=1$, the critical point $\omega_{\rm cr}^1$
is close to but slightly below the maximal value of the mass of 
the ordinary boson stars.
Since the mass of the scalarized boson stars decreases monotonically
until the critical point $\omega_{\rm cr}^2$ is reached,
and since the same holds for the particle number $Q$,
the ordinary boson stars are stable with respect to scalarization
along their first branch. 
Indeed, for a given value of $Q$ along their first branch 
the mass of the ordinary boson stars is always lower than 
the mass of the scalarized boson stars, when these exist.
For $n=2$ the situation is analogous.

For $n\ge 3$, however, the scalarized boson stars assume
their maximal mass no longer at $\omega_{\rm cr}^1$,
but at a smaller value of $\omega$.
Thus they form a (potentially) stable branch,
starting at the first minimum of the mass
and extending until their global maximum.
Along this branch, the scalarized boson stars
represent the energetically favored solutions.
Thus ordinary boson stars will be unstable with respect
to scalarization in this range of frequencies.

We note, that the maximal mass of the scalarized boson stars
increases with $n$.
However, unlike the case of rapidly rotating neutron stars, 
where the maximal mass reached for scalarized neutron stars significantly
exceeds the maximal mass of ordinary neutron stars, the maximal mass of 
scalarized boson stars does not deviate too strongly from the one
of ordinary boson stars.

Considering the end point of the families of scalarized boson star
solutions, we note that for the larger values of $n$, 
with increasing $n$ the end point gets closer to the end point
of the respective family of ordinary boson star solutions.
In particular, the second branches of the ordinary boson
star solutions shorten with increasing $n$.
For $n=8$, the end points of both ordinary and scalarized
boson star solutions are rather close to each other.

\subsection{Scalarized hairy black holes}

\begin{figure}[p!]
\begin{center}
\vspace{-0.5cm}
\mbox{\hspace{-1.5cm}
\includegraphics[height=.25\textheight, angle =0]{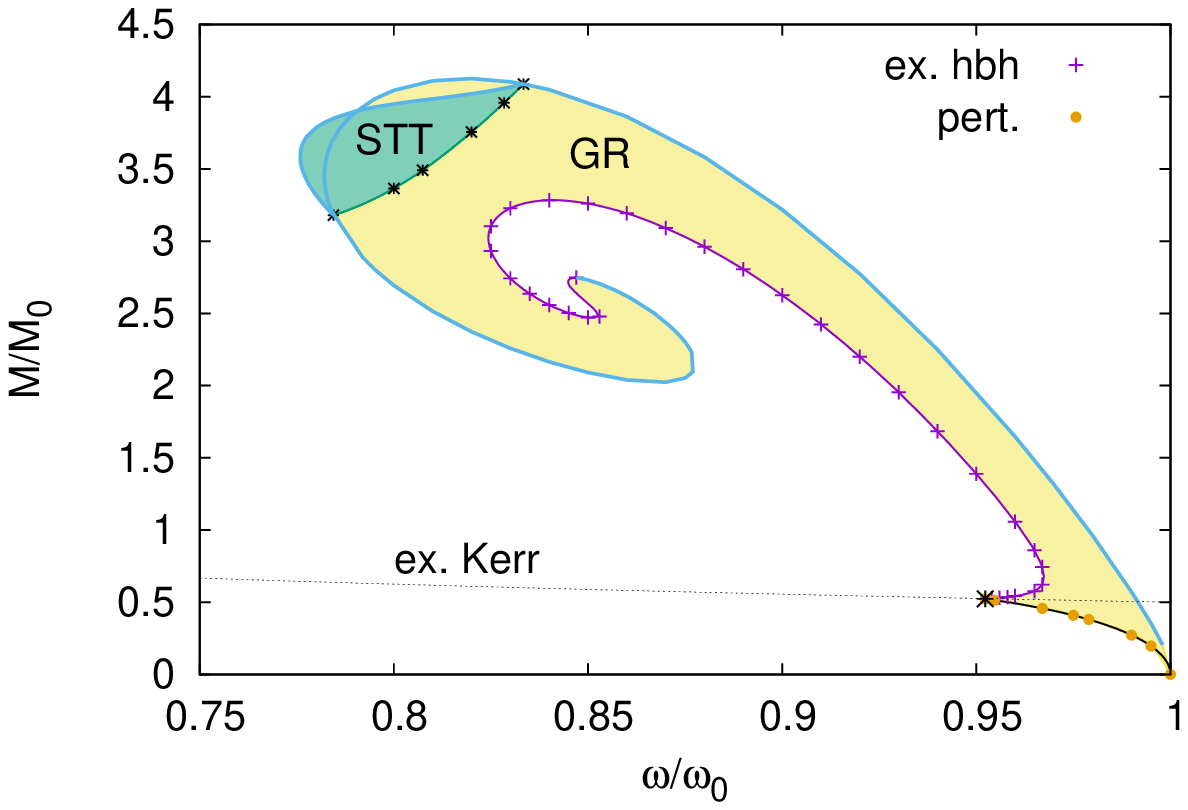}
\includegraphics[height=.25\textheight, angle =0]{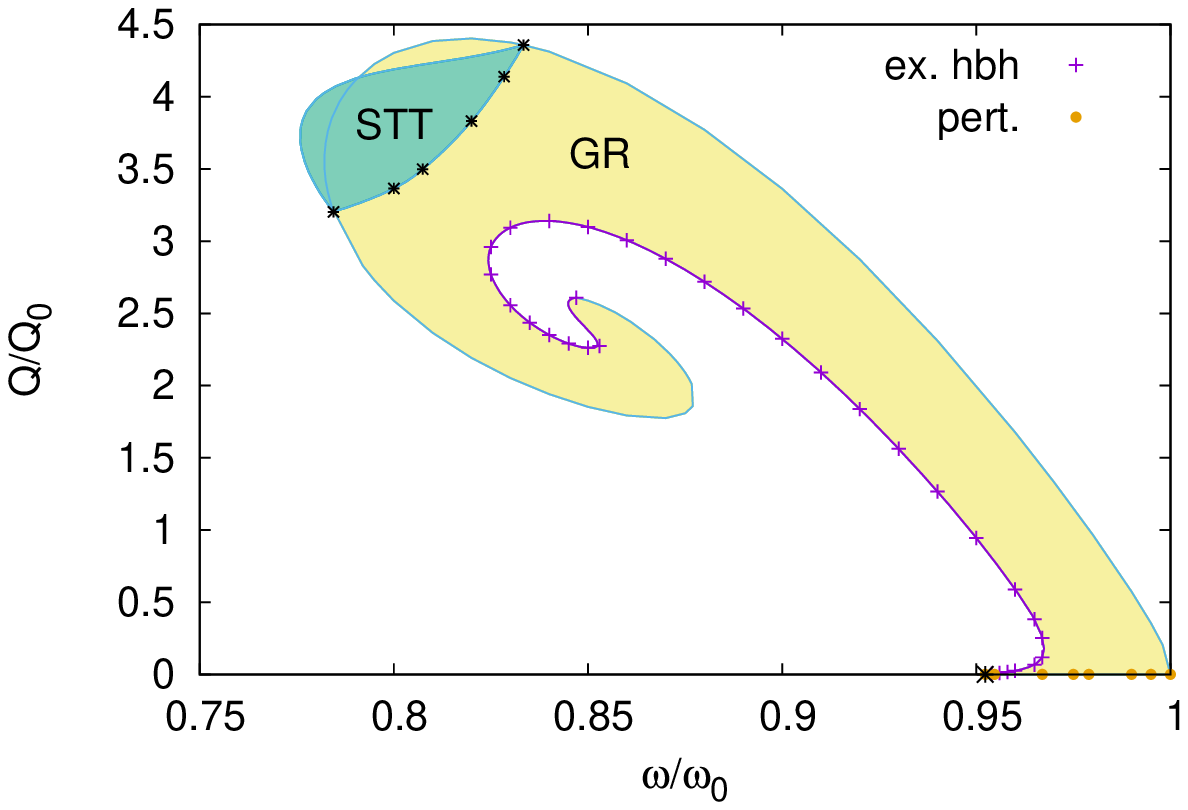}
}
\mbox{\hspace{-1.5cm}
\includegraphics[height=.25\textheight, angle =0]{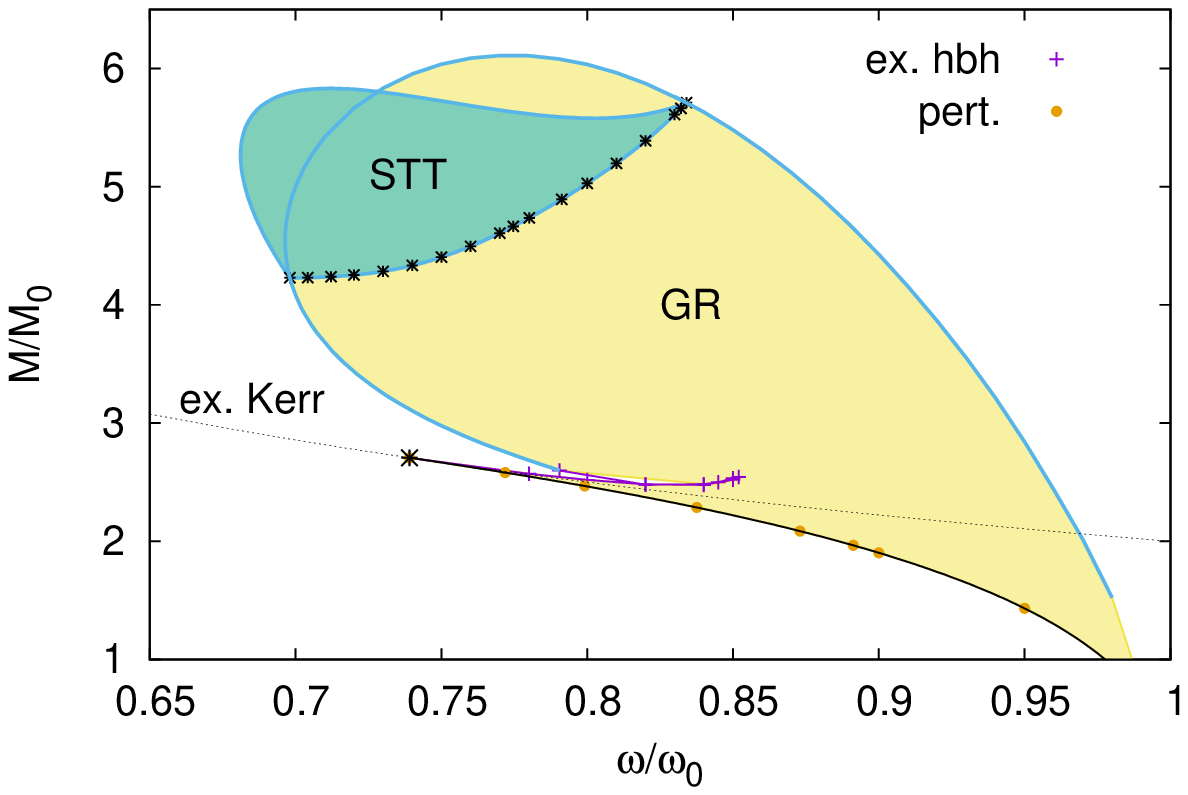}
\includegraphics[height=.25\textheight, angle =0]{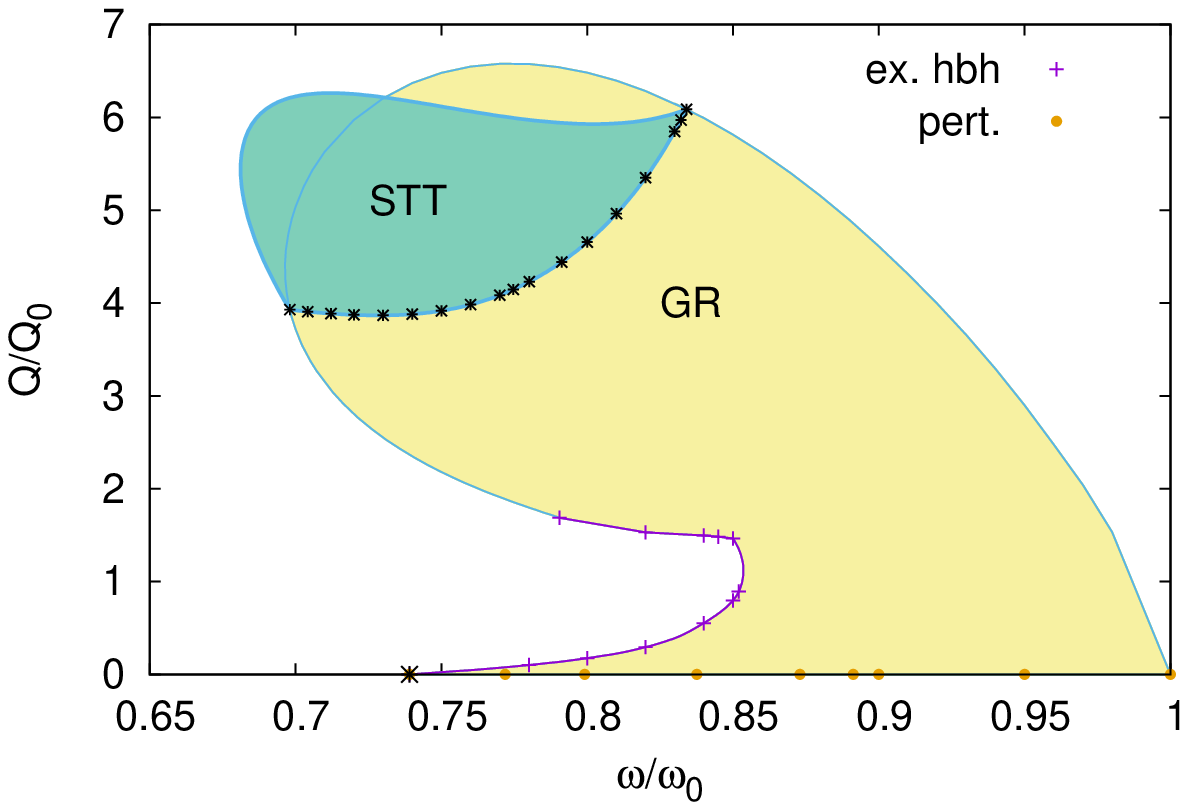}
}
\mbox{\hspace{-1.5cm}
\includegraphics[height=.25\textheight, angle =0]{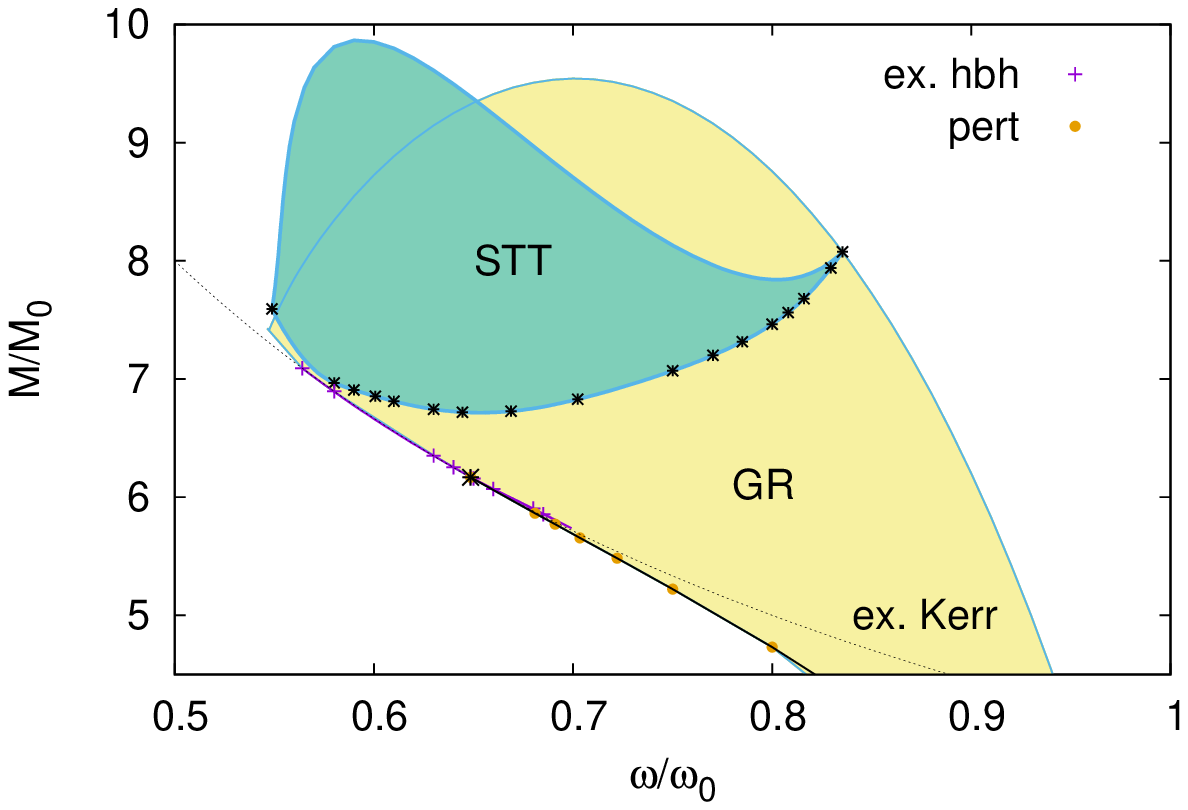}
\includegraphics[height=.25\textheight, angle =0]{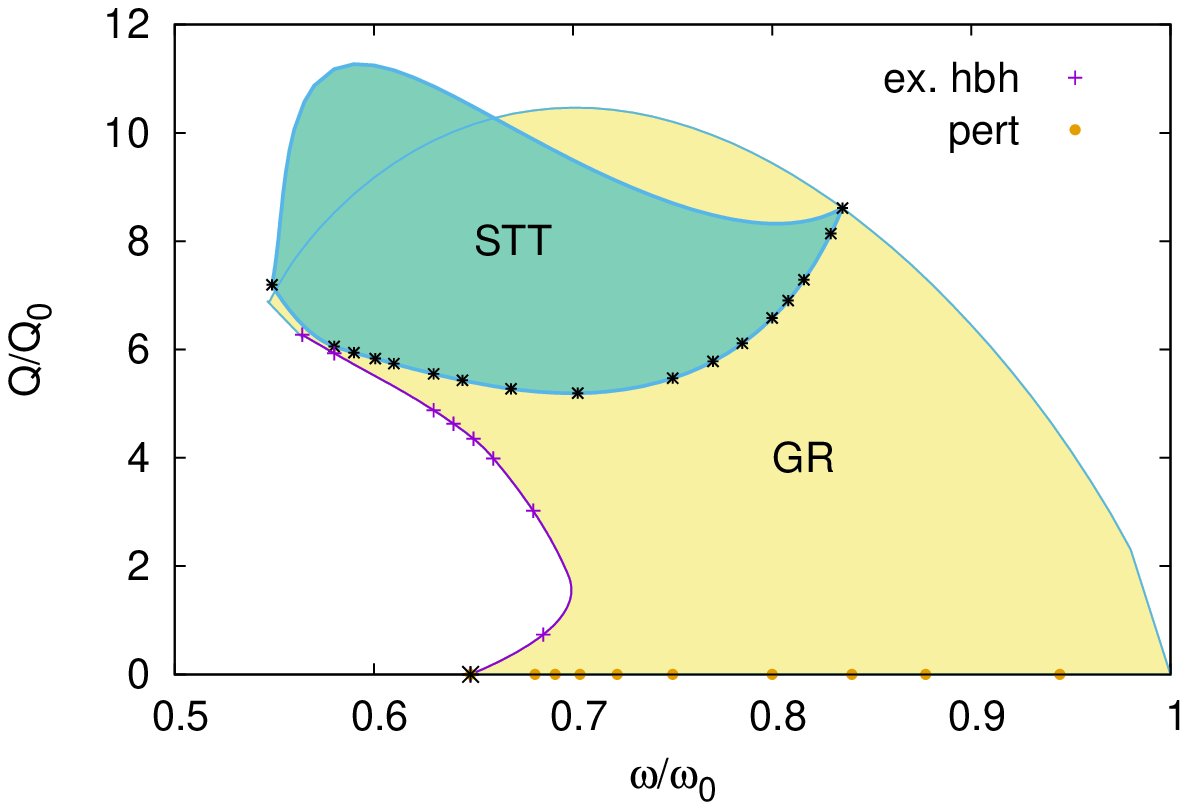}
}
\end{center}
\vspace{-0.5cm}
\caption{
Domains of existence of
scalarized (green) and ordinary (beige) hairy black hole solutions for
rotational quantum numbers $n=1$ (upper), $n=4$ (middle) and $n=8$ (lower).
Shown are the scaled mass $M/M_0$ (left column)
and the scaled particle number $Q/Q_0$ (right column)
versus the scaled boson frequency $\omega/\omega_0$. 
(For black holes $\omega= n\Omega_{\rm H}$.)
The boundaries are formed by boson star solutions, 
extremal hairy black hole solutions,
and perturbative scalar clouds. 
The mass of extremal Kerr solutions is also shown.
\label{fig2}
}
\end{figure}

\begin{figure}[h!]
\begin{center}
\vspace*{-1.5cm}
\mbox{\hspace*{-3.0cm}
\includegraphics[height=.25\textheight, angle =0]{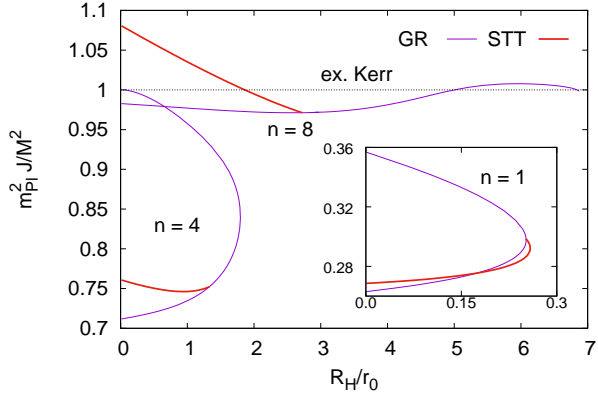}
}
\end{center}

\vspace{-0.5cm}
\caption{The scaled angular momentum $m_{\rm Pl}^2 J/M^2$ is shown versus
the scaled areal horizon radius $R_{\rm H}/r_0$ (\ref{R_H})
for a fixed value of the boson frequency, $\omega/\omega_0=0.8$
for hairy black holes in GR (thin blue) and scalarized hairy black holes (STT) (thick red).
Also shown is the Kerr bound, $ m_{\rm Pl}^2 J/M^2=1$.
\label{fig3}
}
\end{figure}

\begin{figure}[h!]
\begin{center}
\vspace*{-0.5cm}
\mbox{\hspace*{-3.0cm}
\includegraphics[height=.35\textheight, angle =0]{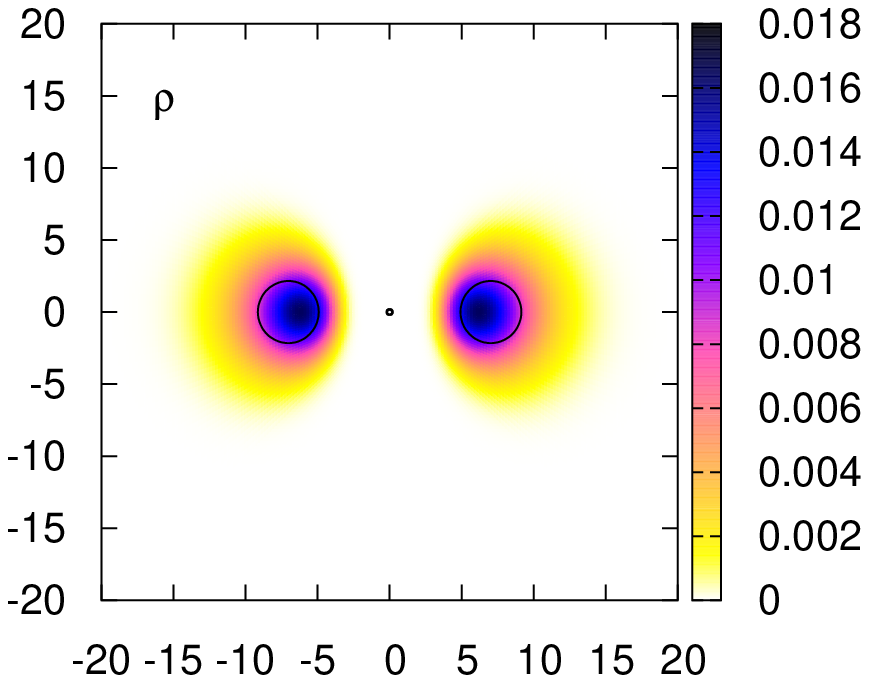}
\hspace*{-3.0cm}
\includegraphics[height=.35\textheight, angle =0]{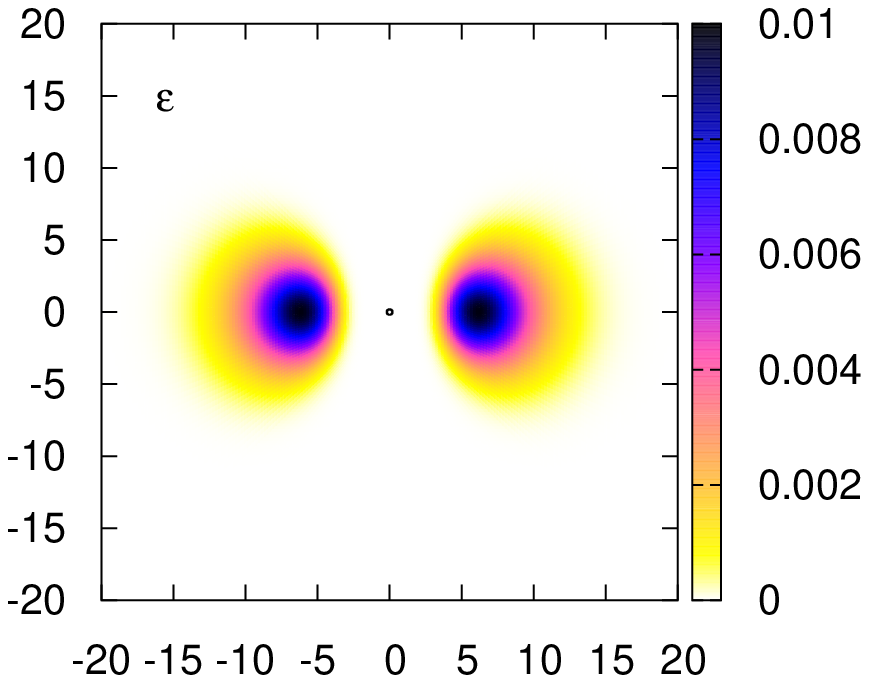}
}

\vspace{-1.5cm}
\mbox{\hspace{-3.0cm}
\includegraphics[height=.35\textheight, angle =0]{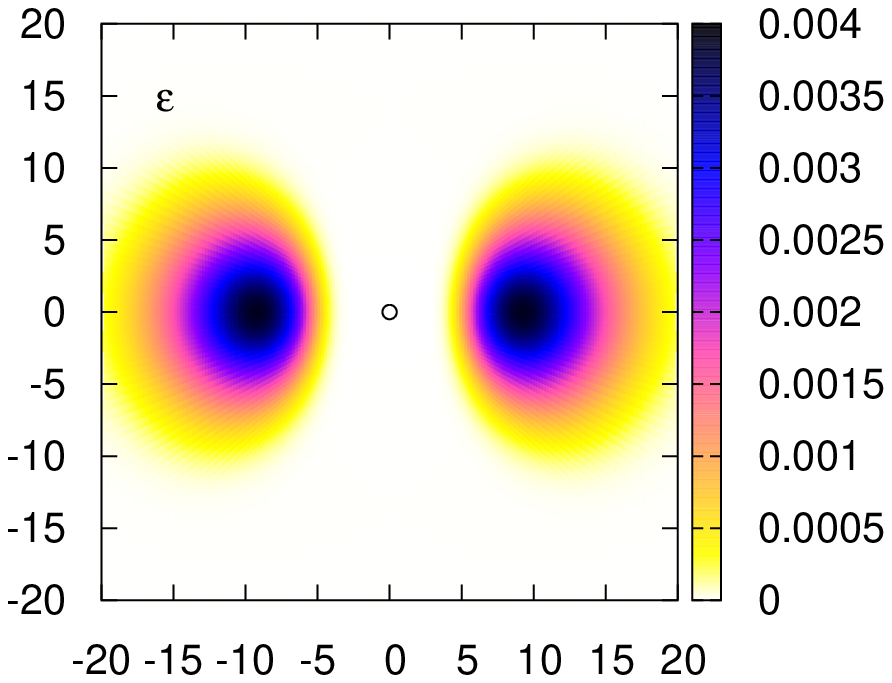}
\hspace{-3.0cm}
\includegraphics[height=.35\textheight, angle =0]{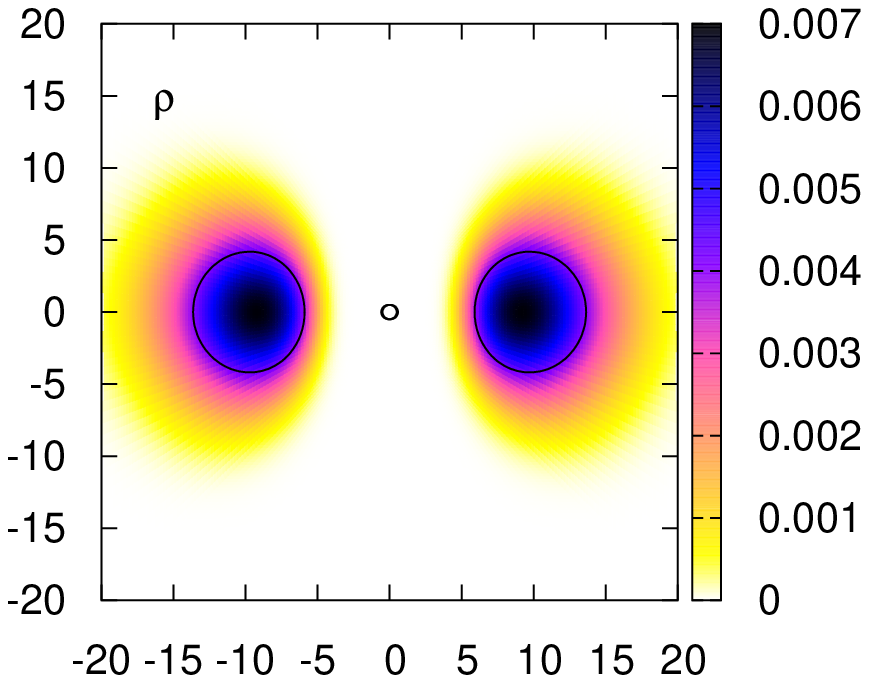}
}
\end{center}

\vspace{-1.5cm}
\caption{Contour plots of the energy-momentum tensor component
$\varepsilon=-T^{\ 0}_0$ (left column) 
and the particle number density $\rho=j^0$ (right column) 
for hairy black holes with rotational quantum number
$n=8$, boson frequency $\omega=0.65\omega_0$,
and mass $M=9 M_0$
in GR (top)
and with scalarization (STT) (bottom).
The axes are $\bar{y} = \pm \bar r \sin \theta$
and $\bar z = \bar r \cos \theta$, both in units of $r_0$.
The solid circles mark the horizon, the dashed lines
the ergosurface.
\label{fig4}
}
\end{figure}

Let us now turn to the hairy black holes.
Starting from a generic ordinary boson star solution, 
a sequence of hairy black holes emerges, 
when the presence of a small horizon is imposed, 
and the horizon is then increased in size.
The domain of existence of hairy black holes is then mapped
by varying the horizon size and the horizon angular velocity.

For ordinary hairy black holes, the domain of existence has been
studied before for $n=1$ and 2 \cite{Herdeiro:2014goa},
employing only a mass term for the boson field.
There it was shown, that the boundary of the domain of existence
of these solutions consists of
\vspace{-0.2cm}
\begin{itemize}
\itemsep=-3pt
\item[(i)] the family of boson stars, 
and the associated families of 
\item[(ii)] extremal hairy black holes and
\item[(iii)] scalar clouds \cite{Hod:2012px,Herdeiro:2014pka}.
\end{itemize}

We exhibit in Fig.~\ref{fig2} the domain of existence of ordinary
hairy black holes for the case of the $\Phi^4$ potential,
and rotational quantum numbers $n=1$, 4 and 8.
In particular, we here show the scaled mass $M/M_0$ (left column)
and the scaled particle number $Q/Q_0$ (right column)
of the hairy black holes versus the scaled boson frequency $\omega/\omega_0$.
We note, that for black holes $\omega= n\Omega_{\rm H}$.
Here we have included the case $n=1$,
to allow for direct comparison with the case
without self-interaction \cite{Herdeiro:2014goa}.
Studies of the rotational quantum numbers $n>3$
were not reported before, neither with nor without self-interaction.
In all these figures, the beige regions labelled GR represent the
domain of existence of the ordinary hairy black holes.

The families of extremal hairy black hole solutions possess
two endpoints. At one endpoint they join the respective branch of
globally regular boson stars solutions in a merger solution.
At the other endpoint they join precisely the respective branch of
scalar cloud solutions. These latter endpoints
are marked in the figures by an asterisk.

In these figures the sets of extremal hairy black hole solutions 
have been obtained by extrapolation, 
where the horizon parameter $\bar{r}_H$
was decreased towards zero.\footnote{We note,
that the extrapolation is no longer completely reliable 
in the innermost region of the spiral-like part for the $n=1$ case.}
The hair of these extremal black hole solutions becomes evident,
for instance, when inspecting the
scaled particle number $Q/Q_0$ of these extremal solutions:
$Q$ is always finite (see Fig.~\ref{fig2}).

As seen in Fig.~\ref{fig2}
for $n=1$, the extremal hairy black hole solutions still form part of a spiral,
whereas for the higher values of $n$ this spiralling pattern is lost.
For $n=4$ and in particular for $n=8$,
the mass of the extremal hairy black hole solutions is close
to the mass of the extremal Kerr black holes. 
The particle number of these extremal hairy black holes is, 
however, clearly finite 
and reaches zero only when the scalar cloud solutions are reached.
The study of $Q$ therefore helps to clarify the domain of existence
and its boundaries.


Let us now consider the domain of existence 
of the scalarized hairy black holes.
It is also exhibited in Fig.~\ref{fig2}
and marked by the green region labelled STT.
The upper boundary of this domain of existence
is always given by the regular scalarized boson stars.
The lower boundary is reached somewhere within the 
domain of existence of ordinary hairy black holes,
at the moment that the scalarization disappears.
Thus we do not observe extremal scalarized hairy black holes.

There is always a part of the domain of existence
of scalarized hairy black holes,
where there are no ordinary hairy black holes.
This part increases with increasing $n$.
For the lower $n$, scalarized black holes exist for smaller
boson frequencies, for the higher $n$
scalarized hairy black holes reach higher values of the mass and the
particle number than ordinary hairy black holes.

Kerr black holes satisfy the bound $m_{\rm Pl}^2 J/M^2 \le 1$.
But this bound may be exceeded by hairy black holes
\cite{Herdeiro:2014goa,Herdeiro:2015gia}.
We demonstrate this in Fig.~\ref{fig3}
for several families of hairy black holes.
In particular, we exhibit the scaled angular momentum $m_{\rm Pl}^2J/M^2$
versus the areal horizon radius $R_{\rm H}/r_0$ (\ref{R_H})
for hairy black holes with $\omega/\omega_0=0.8 $,
$n=4$ and $n=8$ in GR and with scalarization (STT).
We note, that scalarized hairy black holes
can also exceed the Kerr bound.

We exhibit in Fig.~\ref{fig4} contour plots of
the component $\varepsilon=-T^{\ 0}_0$ 
of the energy momentum tensor (left column)
and of the particle number density 
$\rho=j^0$ (right column) of a 
hairy black hole in GR (top) and a scalarized
hairy black hole (bottom).
For comparison, we have chosen 
the same rotational quantum number $n=8$,
boson frequency $\omega=0.65 \omega_0$ and mass $M=9 M_0$
for these black holes.

The central black hole in GR has a horizon area of $A_{\rm H}\approx 0.79 r_0^2$
and is thus much smaller than the scalarized black hole
with $A_{\rm H}\approx 3 r_0^2$.
Both $\varepsilon$ and $\rho$ are concentrated in tori
around the central black hole. The maximal value of $\varepsilon$
is considerably larger for the GR black hole than for the
scalarized one, while the maximal value of $\rho$ is only
slightly larger. 

In the figures showing $\rho$ also the ergosurfaces
are indicated.
Ordinary hairy black holes can feature
ergosurfaces consisting of an ergosphere
and an ergoring, forming together an ergo-Saturn \cite{Herdeiro:2014goa}.
This is also the case for the GR black hole shown.
Here we note, that the same phenomenon may hold as well for
scalarized hairy black holes. Their ergosurface may also 
represent an ergo-Saturn, as depicted in the figure.

\section{Conclusions}

Scalar-tensor theories of gravity offer several observable
consequences (see e.g., \cite{Berti:2015itd}
and references therein). Here we have concentrated on the
effect scalarization.
First, we have shown that scalarization occurs
for rapidly rotating boson stars (with a
fourth order self-interaction).

Rotating boson stars possess a rotational quantum number, 
an integer $n$. Constructing families
of boson stars for $n=1,...,8$, we have shown, that
with increasing $n$, the scalarization becomes more pronounced.
We expect this trend to continue.

Subsequently, we have constructed hairy black holes.
After mapping out the domain of existence of hairy black holes
in GR, which is bounded by boson stars, extremal
hairy black holes and scalar clouds, we have 
surveyed the domain of existence of scalarized black holes.
One boundary of their domain of existence is formed by
scalarized boson stars. 
The other boundary, however, is formed by ordinary hairy black holes.
Here the scalar field simply vanishes,
thus reducing the solutions to
general relativistic solutions with a trivial scalar field.

We have shown that the
physical properties of the scalarized hairy black holes
resemble in many respects those of hairy black holes in GR.
For instance, they may substantially exceed the Kerr bound $m_{\rm Pl}^2 J/M^2 \le 1$,
and they can exhibit ergosurfaces, that consist of two parts,
forming an ergo-Saturn.

The scalarization of rapidly rotating 
boson stars and hairy black holes
allows their mass and particle number to
exceed the maximal values allowed for their general relativistic
counterparts. This effect seen here for the larger values of 
the rotational quantum number 
may be viewed as a downscaled version of what has been
observed for neutron stars.
It should be interesting to increase the rotational quantum
number further, to see how strong the effect of
scalarization may become for rotating hairy black holes.

\vspace{0.5cm}
{\bf Acknowledgment}

\noindent
We gratefully acknowledge support by the DFG within the Research
Training Group 1620 ''Models of Gravity''
and by FP7, Marie Curie Actions, People, 
International Research Staff Exchange Scheme (IRSES-606096).
We gratefully acknowledge discussions with E.Radu.
J.K.~gratefully acknowledges discussions with C.L\"ammerzahl.

\end{document}